%% file: apssamp.tex
\documentclass[reprint, superscriptaddress, amsmath,amssymb, aps, prb, floatfix]{revtex4-2}

\usepackage{graphicx}
\usepackage{dcolumn}
\usepackage{bm}
\usepackage{xcolor}
\usepackage{hyperref}
\usepackage[mathlines]{lineno}
\usepackage{tikz}
\usepackage{orcidlink}
\usepackage{booktabs}
\usepackage[toc,page]{appendix}
\definecolor{lime}{HTML}{A6CE39}

\DeclareRobustCommand{\orcidicon}{
	\begin{tikzpicture}
	\draw[lime, fill=lime] (0,0) 
	circle [radius=0.16] 
	node[white] {{\fontfamily{qag}\selectfont \tiny ID}};
	\draw[white, fill=white] (-0.0625,0.095) 
	circle [radius=0.007];
	\end{tikzpicture}
	\hspace{-2mm}
}

\foreach \x in {A, ..., Z}{\expandafter\xdef\csname orcid\x\endcsname{\noexpand\href{https://orcid.org/\csname orcidauthor\x\endcsname}
			{\noexpand\orcidicon}}
}
\begin{document}
\title{Amplitude-conserving frequency demultiplexer based on a spin wave delay line}
\author{G. Y. Thiancourt\orcidG{}}
\email{gael-yann.thiancourt@universite-paris-saclay.fr}
\thanks{}

\affiliation{%
Université Paris-Saclay, CNRS, Centre de Nanosciences et de Nanotechnologies, 91120, Palaiseau, France
}%
\author{T. Devolder \orcidT{}}
\email{thibaut.devolder@cnrs.fr}
\affiliation{%
Université Paris-Saclay, CNRS, Centre de Nanosciences et de Nanotechnologies, 91120, Palaiseau, France
}%

\date{\today}

\begin{abstract}
Propagating spin waves have emerged as promising candidates for signal processing and information transmission. We present a theoretical investigation of the generation, propagation and collection of spin-wave wave packets in micrometer-wide magnetic conduits, highlighting their potential for analogue frequency demultiplexing in space and time. We clarify the roles of the excitation duration and of the spin wave dispersion relation, as well as the influence of the spin-wave linewidth in setting the spatio-temporal evolution of propagating spin-wave wavepackets excited by Gaussian field pulses. We issue rules for the spatio-temporal separation of wavepackets that carry different frequencies. We finally propose an amplitude-conserving frequency demultiplexer based on a spin wave delay line of 60$\times$5 $\mu$m$^2$ footprint. 

\end{abstract}
\maketitle

\section{Introduction}    
Communication systems require efficient separation of complex multi-tone signals for multichannel processing. This demultiplexing operation can in principle be achieved by exploiting ensembles of frequency carriers ---waves --- that propagate with contrasting group velocities and therefore may separate in space-time. 	
Magnonics is promising platform for this purpose \cite{chumak_magnon_2015}. Spin waves (SWs) are, indeed, ideally suited as they combine microwave frequency operation and $k-$dependent group velocity that can be tuned by material and geometry.
Narrow spin-wave conduits are particularly attractive as confinement effects lead to the appearance of well-defined modes \cite{bayer_spin_2004} with strongly contrasting dispersive characters \cite{demidov_magnonic_2015, perzlmaier_observation_2008}.  Harnessing these modes for spin-wave-based demultiplexing operations requires to understand the dynamics of spin-wave wave packets —an aspect that remains relatively sparsely explored in the community \cite{covington_time_2002,devolder_measuring_2021, davidkova_spin_2025, divinskiy_dispersionless_2021,thiancourt_unidirectional_2024}. \textcolor{black}{The few demultiplexing operations harnessing SWs implemented so far either required an active current supply \cite{martyshkin_magnonic_2024} or used complex geometries \cite{heussner_experimental_2020,zenbaa_universal_2025}.}
	
In this work, we investigate the evolution of SW wave packets upon their propagation. We evidence that wave packets carrying different frequencies separate linearly with the travelled distance $r$ while they broaden only as $\sqrt{r}$ at large $r$, such that they can be separated by propagation. This is the key property that we shall use for demultiplexing purposes. 
\textcolor{black}{Our approach uses the frequency dependence of the group velocity in a single waveguide, so that the frequencies are separated in time and position by a \textit{passive and easy to fabricate} device without requiring complex structures or control currents.}

This paper is organized as follows. After describing the system and the dispersion relations of its relevant SW modes, we compute the frequency dependence of SW-based energy transmission between distant antennas and the corresponding impulse response. Next, we study the spatio-temporal evolution of SW wavepackets of central frequency $\omega_0$ and analyse the respective influences of the group delay (\(\tau_g\)), the group-delay dispersion (\(\tau_D\)) and the excitation duration (\(\tau_0\)). We finally show that frequency demultiplexing can be obtained by a proper engineering of the SW conduit and of the antennas. 

\begin{figure*}
\centering
\includegraphics[width=1.02\textwidth]{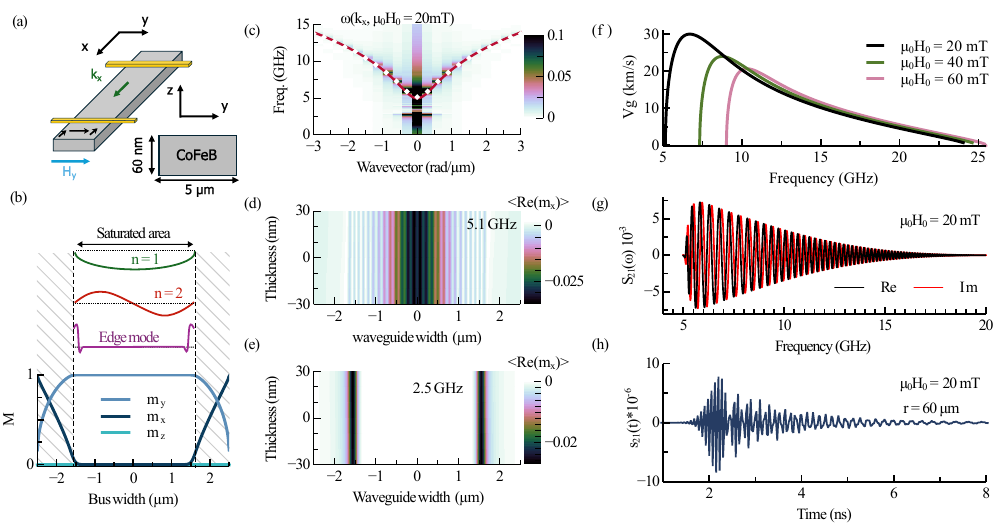}
\caption{Spin-wave properties and response: simulation and analytical modelling of a 5 $\mu$m wide waveguide of  60-nm-thick CoFeB. (a) Sketch of the setup of propagating spin wave spectroscopy (PSWS, \cite{bailleul_propagating_2003}) (b) Configuration of the equilibrium magnetization (bottom-half) and dynamic  magnetizations of the lowest frequency SWs (top-half) for an applied field of 35 mT (c) Map of SWs normalized susceptibility excited by a 500 nm wide antenna RF field. 
The white circles mark the positions of maximum spectral intensity. The red line shows a fit of the analytical spin-wave dispersion relation for the fundamental mode ($n = 1$). (d) and (e) are the simulated mode profiles of width mode and edges modes respectively. 
(f) Analytical group velocity of SWs under various internal fields \(H_0\). (g) Numerical calculation of PSWS response $\tilde S_\textrm{21}(\omega, r=60\,\mu \mathrm{m})$ excited by a 100 nm width antenna. The internal field is $\mu_0H_0=20$ mT. (h) Impulse response corresponding to the previous $\tilde S_\textrm{21}(\omega, r=60\,\mu \mathrm{m})$ of Fig.\ref{fig:dispersion_relation}.g.}
\label{fig:dispersion_relation}
\end{figure*}

\section{Material and geometry} \label{sample}

We harness the propagation of SW wavepackets along a straight magnetic conduit of thickness $t_\textrm{mag}=60$ nm and width $w_\textrm{mag}=5~\mu$m (Fig.~\ref{fig:dispersion_relation}.a) made from a CoFeB alloy of magnetization $M_\textrm{s}=1340$ kA.m$^{-1}$ and damping $\alpha = 0.004$. \textcolor{black}{CoFeB is particularly well suited for device implementation as its high Curie temperature ensures robustness of its properties against temperature variations.}
We apply a field $\mu_0 H_y =35$ mT in the width direction (i.e. $y$), leading the magnetization configuration depicted in Fig.~\ref{fig:dispersion_relation}.b: The magnetization is aligned with the field in the 3~$\mu$m-wide central region of the conduit but the edges remain unsaturated. \textcolor{black}{SEE FOOTNOTE AND ADD REF}\footnote {\textcolor{black}{For implementation in an integrated device, this unpractical applied field of 35 mT could be replaced by two exchange-biasing layers sandwiching the magnetic layer, with the exchange bias set by field annealing in the direction transverse to the magnetic stripe. A structure of the form IrMn (20 nm)/CoFe (2 nm)/CoFeB (56 nm) /CoFe (2 nm) /IrMn (20 nm), with exchange coupling energy of $J=1.3$ mJ/m$^2$ per interface \cite{blachowicz_exchange_2021} would indeed provide an exchange bias whose thickness-averaged effective field amounts to $2 \times J/ (M_\textrm{s} t_\textrm{mag}) = 32~\textrm{mT}$.}}.
Above the conduit, we consider two identical antennas separated by a center-to-center distance $r$, meant to excite and detect SWs with $\vec{k}_x$ the component of their SW wavevectors along the conduit. We thus consider the so-called "Damon–Eshbach" configuration \cite{damon_magnetostatic_1961}. In addition to the propagating character along the conduit length, its finite width is known to lead to some lateral confinement of the SWs. The SWs exhibit consequently a standing wave character in the $\vec{y}$ direction \cite{bayer_spin_2004}. For $t_\textrm{mag}\ll w_\textrm{mag}$ the standing wave character is well described by counter-propagating waves of wavevectors $k_y\approx \pm \frac{n \pi}{w_\textrm{eff}}$, where $n$ is the number of anti-nodes of the mode and $w_\textrm{eff}$ is an effective width corresponding to the saturated area (Fig.~\ref{fig:dispersion_relation}.b). 
   
\section{Spin waves properties in the frequency domain }
\subsection{Exact dispersion relation}
The dispersion relations $\omega(k_x)$ of the set of SWs will be the cornerstone of our study. We first calculate them from micromagnetic simulations using the TetraX package \cite{korber_finite_2021}. Fig.~\ref{fig:dispersion_relation}(c) reports the normalized microwave absorption of the conduit when submitted to the microwave field generated by a 500 nm wide antenna placed at 100 nm above the SW conduit. 
The microwave absorption mostly arises from two modes
: a mode at $2.5$ GHz with a flat dispersion relation and a high frequency one at $5.1$ GHz with a $\vee$-shaped dispersion relation.
    
Their spatial profiles for $k_x=0$ are depicted in Fig.~\ref{fig:dispersion_relation}(d,e), which reports cross sections of the dynamical magnetization component $m_x$. 
The high-frequency mode [Fig.~\ref{fig:dispersion_relation}(d)] has the profile expected for a width mode indexed with $n=1$. In contrast the low-frequency mode is an edge mode, with amplitude being finite solely in two symmetric narrow regions within the edge domains. [Fig.~\ref{fig:dispersion_relation}.e]. While this edge mode is excitable by the antenna, its flat dispersion results in a low group velocity such that this mode cannot transport energy over long distances. From now on, we shall therefore only consider the high-frequency (i.e. $n=1$) mode
. As it is convenient to have an analytical description of the dispersion of this mode, we now compare the results of TetraX to the approximate expressions of Kalinikos and Slavin \cite{kalinikos_theory_1986}.

\subsection{Approximative analytical dispersion relation}
For a transversally magnetized stripe, the dispersion relation of the $n=1$ mode can be approximated by \cite{demidov_magnonic_2015}:
    \begin{equation}
    	\omega(k_x) =
    	\gamma_0
    	\sqrt{
    	\big(H_0 + M_s (1 - P)\big)
    	\left(H_0 + M_s P \frac{k_x^2}{k^2}\right)
    	},
    	\label{wk}
    \end{equation}
    where $P = 1 + \frac{e^{-k t_{\mathrm{mag}}} - 1}{k t_{\mathrm{mag}}}$, $k^2 = k_x^2 + \pi^2 / w_{\mathrm{eff}}^2$, \textcolor{black}{$\gamma_0 / (2 \pi)=29.3$ GHz/T the gyromagnetic ratio} and $H_0$ accounts for the adequately averaged internal magnetic field. The effective demagnetizing correction can be found using a fit [red line in Fig.~\ref{fig:dispersion_relation}(c)] of the TetraX dispersion relation with Eq.~\ref{wk}, which yields $\mu_0 H_0 = 20$~mT  $< \mu_0 H_y = 35$~mT. The frequency linewidth $\Delta \omega_0$ is: (see Supplemental Material, section \ref{app:deltaomega0}),
	\begin{equation}
		\Delta\omega_0(k_x,k_y)=\alpha\gamma_0\left(2H_0+M_s(1-\frac{k_y^2}{k_x^2 +k_y^2 }P)\right). 
		\label{DeltaOmega}
	\end{equation} 
Since $M_s \gg H_0, P M_s$ the linewidth is almost wavevector independent. We will consider it to be constant.

The approximate dispersion relation (Eq.~\ref{wk}) was confirmed to work in the ranges of wavevectors and fields to be used in this paper. Fig.~\ref{fig:dispersion_relation}(f) displays the corresponding field and frequency dependences of the group velocity $v_g$. Increasing the field increases the frequency gap while essentially preserving the overall SW propagation metrics. The $v_g(f)$ curves differ only at low frequencies where the $v_g$'s systematically peak, reaching for instance $30~\mathrm{km/s}$ for $\mu_0H_0 = 20$~mT at the inflexion point of the dispersion relation at 7~GHz. We will use the strong variation of the $v_g(f)$ with the frequency for the space-time separation of wavepackets.

\subsection{Characteristic times and lengths of spin wave propagation}
Prior to modelling the propagation of SW wavepackets, it is worth defining the three characteristic times involved in the problem. \\
(i) The first one is the initial duration $\tau_0$ of a wavepacket of central frequency $\omega_0/(2\pi)$, to be defined in section \ref{DefiningGaussianPulse}.  

The two other characteristic times are defined by expanding the dispersion relation and considering the propagation of a SW over a distance $r \geq 0$.\\ 
(ii) The arrival time $\tau_g$ of the wavepacket is the group delay: 
\begin{equation}
    \tau_g(\omega_0,r) = r\,\frac{d k_x(\omega)}{d\omega}\bigg|_{\omega=\omega_0}=\frac{r}{v_g(\omega_0)},
    \end{equation}
    which expresses the spatio-temporal drift of the centre of the wavepacket at the group velocity $v_g(\omega_0)$.
\\  
(iii) The third characteristic time of the problem is the group-delay dispersion $\tau_D(\omega_0,r)$ (GDD) defined from the curvature of the dispersion relation:
\begin{equation}
    \tau_D^2(\omega_0,r) = r\,\frac{d^2 k_x(\omega)}{d\omega^2}\bigg|_{\omega=\omega_0}.
    \label{DefinitionGDD}
\end{equation}
$\tau_D^2$ characterizes the progressive evolution of a wavepacket upon its spatio-temporal propagation. The first meaning of $\tau_D^2$ can be understood from dimensional analysis: it is the square of a time, or equivalently, a time divided by a frequency. Qualitatively, $|\tau_D^2|$ parametrises the increase of a pulse duration (typically in added nanoseconds) that a pulse of given frequency width (typically in GHz) undergoes after having propagated over a distance $r$. The second meaning of $\tau_D^2$ resides in its  sign. 
It is negative (positive) when the high-frequency components propagate faster (slower) than low-frequency components.  As a result, the wavepackets progressively distorts (it "chirps"). For $\tau_D^2 <0$ the low frequency components progressively gather near the trailing edge of the wavepacket, while the high frequency ones group at the leading edge of the wavepacket. 

\textcolor{black}{
  The interest of defining $\tau_g$ and $\tau_D$ is evident when considering a narrow band wavepacket of central frequency $\omega_0$ and central wavevector $k_0(\omega_0)$. When travelling a distance $r$, each frequency component of this wavepacket acquires a phase shift $\phi(\omega)=k~r$, that can be expressed to second order in $\delta\omega=\omega-\omega_0$ as: 
    \begin{equation}
        \phi(\omega)= \phi_0
        + \tau_g(\omega_0,r)\,\delta\omega
        + \tfrac{1}{2}\tau_D^2(\omega_0,r)\,\delta\omega^2
        + \mathcal{O}(\delta\omega^3).
        \label{eq:Phi_expand}
    \end{equation}}
The last important metrics is the attenuation length of the SWs $ L_{\mathrm{att}} = \frac{2 v_g}{\Delta \omega_0}$ \cite{gladii_spin_2016}. 

It is worth giving order-of-magnitudes, for instance $r = 60~\mu\mathrm{m}$ and $\mu_0H_0=20$ mT. A wavepacket central frequency $\omega_0/2\pi = 10~\mathrm{GHz}$ yields $\tau_g = 3~\mathrm{ns}$ and $\tau_D^2 = (0.26~\mathrm{ns})^2$. Comparatively, the SWs are faster for $\omega_0/2\pi = 6.6~\mathrm{GHz}$ since the arrival time is decreased to $\tau_g = 2~\mathrm{ns}$. Besides the SWs are substantially less dispersive at this frequency, the dispersion delay is strongly reduced to $\tau_D^2 = - (0.08\,~\mathrm{ns})^2$.
The mode linewidth is 0.19 GHz. This leads to an attenuation length of $48~\mu$m at $k_x=$0.5 rad/$\mu$m and $5.5~\mu$m at 10 rad/$\mu$m. 


\subsection{Antenna-to-antenna transmission spectrum}	
Let us now calculate the transmission $\tilde{S}_{21}(\omega, r)$ between two antennas. The transmission refers to the complex inductive voltage induced in the receiving antenna (P2) when the input antenna (P1) is driven by an RF current at frequency $\omega$. Following Eq.~(18) of Ref.~\cite{devolder_propagating-spin-wave_2023} and considering only the in-plane component of the antenna field, it is:
\begin{equation}
		\tilde S_{21}(\omega,r)=\int_{-\infty}^{-\infty}\chi^{xx}(k_x,\omega)e^{-ik_x r}\tilde h_{rf}(k_x)^2\,dk_x,
		\label{eq: S21}
\end{equation} 
where $\tilde h_{rf}(k_x)^2=\tilde h^{P1}_{rf}(k_x)\tilde h^{P2,*}_{rf}(k_x)$ is the product of the efficiency functions of antenna P1 and the complex conjugate of that of antenna P2, describing the SW wavevectors that can couple inductively to the antennas (See Eq.~6 of ref.~\cite{devolder_propagating-spin-wave_2023}). For a thin single (S) wire antenna of width $L_1$ directly deposited above the magnetic material, the antenna function is the real-valued function:
\begin{equation}
    h_{rf}^S(k_x) \propto \textrm{sinc} \left(\frac{kL_1}{2}\right).
\end{equation}
$\chi^{xx}(k_x,\omega)$ is the transverse susceptibility. In the low damping regime where $\alpha\ll1$ it is (see Supplemental Material, section~\ref{app:deltaomega0}):
	\begin{equation}
	\tilde\chi^{xx}(k_x,\omega)=\frac{\omega_\textrm{max}(k_x)^2}{\omega_k(k_x)^2-\omega^2-i\omega\,\Delta\omega (k_x)},
	\end{equation} where $\omega_k(k_x)$ is the SW dispersion relation, $\Delta\omega (k_x)$ the SW linewidth and $\omega_\textrm{max}(k)^2$ is $\gamma_0^2M_s(H_0+M_s(1-P)$. An example of susceptibility spectrum is plotted in Fig.~\ref{fig:Excitation}.c. 
    
Fig.~\ref{fig:dispersion_relation}.g. shows the calculated \footnote{To save time the transmission coefficient for negative frequencies can be constructed from the Hermitian symmetry condition $\tilde S_{21}(-\omega,r)=\tilde S_{21}^*(\omega,r)$, which reflects the real-valued nature of the voltage when in the time domain. The integration is performed over $k \in [-20,20]~\mathrm{rad}\,\mu\mathrm{m}^{-1}$ with a step size of $0.01~\mathrm{rad}\,\mu\mathrm{m}^{-1}$, and over a frequency range from 4 to 20~GHz with a resolution of 2~MHz (The response is assumed null below 4 GHz). Caution must be taken not to alias the frequency spectrum by using an insufficiently large frequency step.} spectrum of the antenna-to-antenna transmission for $\mu_0H_y=20$ mT, 100 nm-wide antennas and $r=60~\mu$m. 
The transmission is allowed above the SW gap at 5.1 GHz and consists of a phase-rotating factor with an envelope that decreases with frequency and extends up to circa 16 GHz.

The transmission response to any stimulus of spectrum $\tilde G(\omega)$ can be obtained by multiplying it with the transfer function: 
	\begin{equation}
		\tilde S_\textrm{out}(\omega,r)=\tilde S_{21}(\omega,r)\tilde G(\omega).
		\label{eq:Eout}
	\end{equation} 
The corresponding time-domain response is the convolution of the input signal with the system's impulse response, or equivalently the Fourier transform of Eq.~\ref{eq:Eout}. 

\section{Spin waves properties in the time domain}	
\subsection{Antenna-to-antenna impulse response}	
$\tilde S_{21}(\omega,r)$ can be transformed to describe the impulse response, i.e. the time-domain response at the output antenna to a Dirac field pulse applied at the input antenna \cite{devolder_measuring_2021}. Fig.~\ref{fig:dispersion_relation}.g displays such impulse response. As expected for $r$ = 60 $\mu$m and $v_g^\textrm{max} \approx$ 30 km/s the first SWs arrive at $r/v_g \approx$ 2 ns. The contribution from the slower SWs is present at larger arrival times, typically up to 6 ns. This distribution of arrival times stems from the frequency dependence of group velocity. This opens the possibility to separate in time and space the different frequency contributions inside a complex signal. As a first case study, let us analyse the response to a short burst of SWs.


\subsection{Defining a Gaussian pulse} 
\label{DefiningGaussianPulse}


    \begin{figure}
        \centering
        \includegraphics[width=1\linewidth]{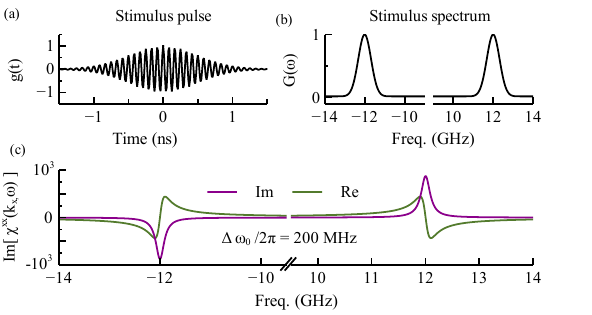}
            \caption{Applied microwave field pulse in time- and frequency-domains, and magnetic susceptibility. (a) 12~GHz Gaussian pulse of duration of $\tau_0 = 0.5$~ns or equivalently a full width at half maximum $\tau_0^\textrm{FWHM} = 1.17$~ns. (b)  Frequency spectrum of the excitation. (c) Magnetic susceptibility for mode linewidths $\Delta\omega_0 = (2\pi)\times 200$~MHz and wavevector $k_0=2\,\textrm{rad}/\mu$m. }
        \label{fig:Excitation}
    \end{figure}

We consider a pulsed input field $g(t)$ of Gaussian envelope, amplitude $E_0$, root mean square (RMS) duration $\tau_0$, which means a pulse "duration" (full width at half maximum of the envelope) being $\tau_0^\textrm{FWHM}\approx2.35\,\tau_0$. It reads:
\begin{equation}
		g(t) = E_0\,e^{\!-\tfrac{t^2}{2\tau_0^2}} \textrm{cos}(\omega_0t).       
        \label{eq:excitation}
    \end{equation}
The associated spectrum $\tilde{G}(\omega)$ is real-valued consists of two Gaussian functions centred at $\pm\omega_0$:
	\begin{equation}
		\tilde{G}(\omega) = \frac{E_0\tau_0}{2}\left[\,e^{\!\tfrac{-
        \tau_0^2}{2}(\omega 	- \omega_0)^2}+\,e^{\!\tfrac{-\tau_0^2}{2}(\omega 	+ \omega_0)^2}\right] \in \mathbb R.
	\label{eq:Gauss}
	\end{equation}  

    Fig.~\ref{fig:Excitation}.a and b respectively display the excitation pulse in time and frequency domains for $f_0=\omega_0(2\pi)^{-1}=12$ GHz and a pulse duration $\tau_0^\textrm{FWHM}=1.2$ ns.

\section{Analytical approximate model for pulsed operation} \label{SectionAnalaytical}
Let us calculate the response of the SW system to this Gaussian pulse. An analytical approximation can be derived under the assumptions that \\(i) the dispersion relation is a parabola \footnote{This assumption restricts the applicability of the model to situations where the pulse duration follows $\tau_0 \gg \tau_D$ so that the frequency content of the input pulse is well within the parabolic part of the dispersion relation.}, \\(ii) the antenna is broadband in wavevector space, \\(iii) the material only responds to the in-plane (i.e. $x$-) component of the antenna field \footnote{A complete description of the transmission signal $\tilde S_{21}(\omega,r)$ requires accounting for all components of the rf magnetic field generated by the antenna. This can be achieved by summing the contributions arising from the in-plane and out-of-plane rf-field components$(h_{\mathrm{rf}}^{x}, h_{\mathrm{rf}}^{z})$, each weighted by the corresponding dynamic susceptibilities}.

\subsection{Efficiency function of the broadband antenna}

All along section \ref{SectionAnalaytical} we shall consider the limit of an antenna of vanishing width and thickness, i.e. $\{w_{\mathrm{ant}},p\}\rightarrow 0$.
 In this limit, the antenna generates an infinitely localized field: 
    \begin{equation}
        h_{\mathrm{rf}}(x,\{w_{\mathrm{ant}},p\rightarrow0\})
        = \delta(x)
        \label{eq:Hrfx}
    \end{equation} where $\delta(x)$ is the Dirac delta function.
    In reciprocal space, the antenna efficiency function is constant:
    \begin{equation}
        h_{\mathrm{rf}}(k,\{w_{\mathrm{ant}},p\rightarrow0\})
        = 1. \label{IdealizedAntenna}
        \end{equation}
In the analytical framework of the present section (\ref{SectionAnalaytical}) we consider that the SW wavepacket is excited by an rf field of the form $g(t) \delta(x)$, i.e. a Gaussian-in-time and infinitely localized at the sole $x=0$ position. This will be compared later to a realistic case. 

\subsection{Space-time evolution of a wavepacket}    
Since the rf pulse (Eq.~\ref{eq:excitation} and \ref{eq:Gauss}) is centred about $\omega_0$ it excites SWs mainly about $k_0$. By expanding the dispersion relation using Eqs.~\ref{eq:Phi_expand} and \ref{eq:Eout}, we obtain a first-order description of the evolution of the wavepacket. At a time $t>0$ and at a distance $r$ from the antenna, the output signal is (see Supplemental Material, section~\ref{app:analytical}):

\begin{equation} s_{\mathrm{out}}^{k_0}(r,t) \propto \overbrace{L(|r|)}^{ \textrm{damping}} ~ \underbrace{B(t, r)}_{\textrm{broadening}}~ \overbrace{C(r)}^\textrm{decoh} ~ \underbrace{\cos[\varphi(r,t)]}_\textrm{oscillations} \label{eq:e(t)atk0}
\end{equation} where the first factor is a damping-induced propagation loss, the second factor  describes the dispersion-induced broadening of the wavepacket, the third factor describes a dispersion-induced internal decoherence of the wavepacket, and the last factor describe the propagation of the phase front. Let us discuss these factors.

\subsubsection{Amplitude loss due to Gilbert damping}

The first factor in Eq.~\ref{eq:e(t)atk0} is the classical propagation loss arising from the Gilbert damping, expressed as:
\begin{equation}
{L(|r|)} = e^{-|r|/L_{\textrm{att}}^0}
\end{equation} with $L_{\textrm{att}}^0 = \frac{2 v_g}{\Delta \omega_0}$ the attenuation length of the SW $\{k_0, \omega_0\}$. 

\subsubsection{Dispersion-induced broadening of the wavepacket}
The second factor in Eq.~\ref{eq:e(t)atk0} is: 
\begin{equation} B(t, r)= \exp\left({-\frac{\tau_0^2\,}{2\tau_{\mathrm{out}}^4}} \left(t-\frac r {v_g^0}\right)^2 \right)
\label{eq:GaussianTerm}
\end{equation}
It describes the temporal broadening of the wavepacket about its centre which drifts in time-space as $r=v_g^0 t$ according to the group delay. The temporal width of the wavepacket at a propagation distance $r$ is $\tau_{\mathrm{out}}^2 / \tau_0$ with: 
\begin{equation}
    \tau_{\mathrm{out}}^4(r)=\tau_0^4+\tau_D^4(r) ~~~\in \mathbb{R}^+.
\label{DefinitionTauOut}\end{equation}

\subsubsection{Wavepacket decoherence}
 In Eq.~\ref{eq:e(t)atk0} the temporal width $\tau_{\mathrm{out}}^4(r)$ also affects the decoherence factor $C(r)$. $C(r)$ encodes how the amplitude of the wavepacket is decreased by destructive interferences between its different spectral components. It reads:
   \begin{equation}
    C(r)
    =\frac{E_0\pi\,\tau_0}{[\tau_{\mathrm{out}}^4(r)]^{1/4}}\,
    \frac{\omega_{\mathrm{max}}^2}{\omega_0v_g}.
    \end{equation}

\subsubsection{Phase oscillation within the pulse envelope} 
Finally, the phasor term $\cos[\varphi(r,t)]$ captures both the wave character of the carrier and the dispersion-induced phase chirp. It is given by:
\begin{equation}
\varphi(r,t) =
\underbrace{(k_0 r - \omega_0 t)}_{\text{carrier wave}} \;+\;
\underbrace{f(t, \tau_0, \tau_{\mathrm{vg}}, \tau_D^2)}_{\text{chirp-induced phase}}
\label{eq:phase}
\end{equation}
The carrier term describes the wave character, with a phase front advancing at $r=\omega_0 t / k_0$. The other contribution has a marginal quantitative influence. It is discussed in the supplemental material. Overall, the shape of the wave-packet results from the interplay between the dispersion, the loss and the finite excitation duration, a behaviour that is characteristic of wave propagation in dispersive media \cite{agrawal_fiber_2012}. 
        
\subsection{Physically-instructive limits} 
It is instructive to examine two specific limits of Eq.~\ref{eq:e(t)atk0}: the time evolution right under the antenna, and the wave packet profile at time zero. 

\subsubsection{Time evolution of the wavepacket right below the input antenna} 
If the receiving antenna was at the same place as the input antenna (i.e. at $r=0$), the group-delay dispersion would vanish and Eq.~\ref{eq:e(t)atk0} would reduce to
\begin{equation}
s_{\mathrm{out}}^{k_0}(t,r=0)
\propto -\frac{E_0\pi  e^{-\frac{t^2}{2 \tau_0^2}} }{v_g \omega_0}\sin (\omega_0 t).
\label{SanityCheck1}
\end{equation}
This resembles the excitation pulse of Eq.~\ref{eq:excitation}: it lasts $\tau_0$ and carries a harmonic oscillation. This sinus-like response is in quadrature with the cosinus-like stimulus $g(t)$. The quadrature relationship is as expected for a precessional response of the magnetization to a field.

One may be surprised not to find a factor like $e^{-\Delta \omega_0 \, t /2 }$ that would express some damping of the SWs within Eq.~\ref{SanityCheck1}. This apparent absence of the role at damping is related to the \textit{propagating wave} situation that suppresses any \textit{local} memory. Indeed at any given time the SWs excited in the past at $r=0$ have propagated away and are not longer at this central position, such that their damping ---though real--- is irrelevant there.

\subsubsection{Profile of the wavepacket at the time origin} 
The second instructive limit is the spatial profile of the wavepacket at $t=0$, which reads
\begin{equation} \label{riri}
s_{\mathrm{out}}^{k_0}(t=0,r)
=  C(r)
e^{-\frac{\lvert r \rvert}{L_{\mathrm{att}}}}
e^{-\frac{r^2}{2\tau_{\mathrm{out}}^2\,v_g^2}}\cos[\varphi(r,0)]
,
\end{equation}
It is noticeable that while the excitation field pulse was infinitely localized (Eq.~\ref{eq:Hrfx}), the response of the magnetization is \textit{not} localized and has a finite spatial extension, a counter-intuitive result that deserves explanation. 

This finite extension at $t=0$ results in fact from the propagation of the SWs that were generated \textit{in the past} (recall that the stimulus pulse starts essentially at $t \approx -\tau_0$, see Fig.~\ref{fig:Excitation}a). At $t=0$ these past SWs have propagated over a typical distance $v_g \tau_0$ and, meanwhile, they have undergone a dispersion-induced broadening set by $v_g \tau_D$, both effects giving a finite spatial extension. These propagation and dispersion effects affecting the SWs created in the past are gathered in the $\tau_\textrm{out}$ (see its expression in eq.~\ref{DefinitionTauOut}) contribution inside eq.~\ref{riri}. Meanwhile, all these past SWs have suffered a dissipation-induced decay that can be seen in the $\textrm{exp}[|r|/L_\textrm{att}]$ factor.

We note that the eqs.~\ref{SanityCheck1} and \ref{riri} are singular when the dispersion relation is flat (i.e. $v_g=0$). This pathological situation is treated in the Supplemental Material.

\begin{figure}
	\includegraphics[width=0.5\textwidth]{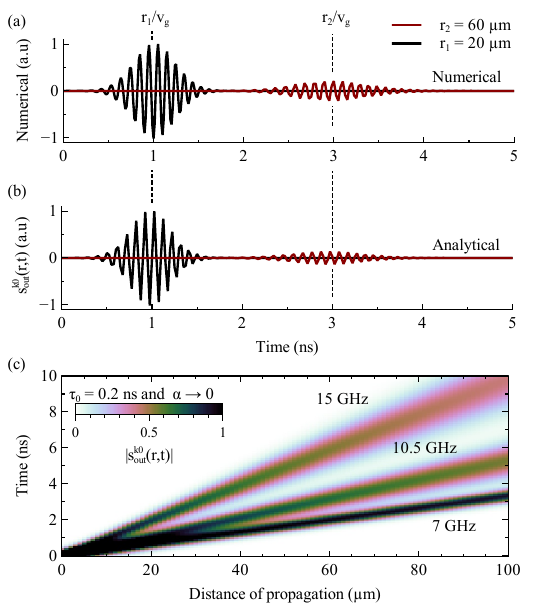}
    \caption{Voltages expected at the receiving antenna arising for stimuli of durations $\tau_0 = 0.2~\mathrm{ns}$. (a): Numerical integration [Eq.~\ref{eq: S21}) and \ref{eq:Eout}]. Parameters: $W_\textrm{stripe}=5~\mu$m, $w_\textrm{ant}=100$ nm, $\mu_0 H_0=$ 20 mT and \(\alpha=0.004\). The stimulus central frequency is 10 GHz. (b): Analytical evaluation using Eq.~\ref{eq:e(t)atk0} for propagation distances \(20 \) and \( 60 ~\mu\mathrm{m} \). (c): Evaluation of $\big|\big|s^{k_0}_{21}(t,r)\big|\big|$ for three wave packets as a function of propagation distances and times. The packets are generated from pulses of central frequencies 7, 10.5 and 15 GHz in the same material as (a) and (b) but with vanishing damping to illustrate the sole effect of dispersion. 
}
	\label{fig:ModelVSAnal}
\end{figure}

\section{Discussion of the time-space evolution of the wave packets }
The space-time evolution of a wavepacket as modelled through Eq.~\ref{eq:e(t)atk0} was derived under three simplifying assumptions that restrict its applicability. In this section we compare this previous analytical formulation with the exact result obtained by numerical integration of Eq.~\ref{eq:Eout} and subsequent Fourier transformation.

\subsection{Comparison between the exact result and the analytical approximation} \label{ComparaisonDesModeles}
The time-resolved transmitted voltages are calculated in Fig.~\ref{fig:ModelVSAnal}(a,b) for antenna-to-antenna distances $r_1 = 20~\mu\mathrm{m}$ and $r_2 = 60~\mu\mathrm{m}$, field pulse duration $\tau_0 = 0.2$~ns, central frequency 10~GHz, stripe width 5 $\mu$m, internal field $\mu_0 H_0 = 20$~mT and mode linewidth $\Delta\omega_0=(2\pi)\times 200~\textrm{MHz}$. The panel (a) is the exact numerical integration [Eq.~\eqref{eq:Eout}] for a 100-nm-wide antenna. The panel (b) is the analytical estimate [Eq.~\ref{eq:e(t)atk0}] that assumes an antenna of zero width \footnote{The curves are evaluated using the values of $\tau_D$ and $\tau_g$ extracted from a parabolic Taylor expansion of the dispersion relation about 10 GHz.In both models the wave packets are normalized to the peak amplitude at the shortest propagation distance ($20~\mu\mathrm{m}$).}.

Overall, the two models agree on the arrival times of the wavepackets, their temporal broadening, their attenuation and their phase oscillations. As expected for a 10 GHz input pulse, the travelling wavepackets exhibit time oscillations of pseudo-period 100 ps [Fig.~\ref{fig:ModelVSAnal}(a,b)]. They arrive at the times that were formerly anticipated from the group delays, i.e. at $t_1 \approx 1$~ns for $r_1 = 20 ~\mu$m and  $t_2 \approx 3$~ns for $r_2 = 60 ~\mu $m. 
\textcolor{black}{The agreement between the analytical and numerical model is almost perfect: the full widths at half maximum (FWHM) of the wavepackets are strictly identical for both datasets. The arrival times also match: they are $\tau_g^\text{num}(20\,\mu\text{m}) = 2.95$~ns and $\tau_g^\text{num}(60\,\mu\text{m}) = 0.98$~ns, and $\tau_g^\text{ana}(20\,\mu\text{m}) = 2.95$~ns and $\tau_g^\text{ana}(60\,\mu\text{m}) = 1.01$~ns.} 

The propagation loss is also visible as an amplitude decay from $r_1$ to $r_2$. The dispersion is clearly perceivable: the wavepackets have broadened compared to the initial width $\tau_0^{FWHM}=0.47$ ns, with FWHM width being 0.54 ns for $r_1 = 20 ~\mu \textrm m$ and 0.96 ns for $r_2 = 60 ~\mu \textrm m$.
\textcolor{black}{This illustrates the contribution of dispersion which increases with the propagation distance. 
We emphasize that a 100 nm wide antenna is small enough not to contribute to the chirp of the wavepacket. Indeed, a SW at 10~GHz, Fig.~\ref{fig:ModelVSAnal}(a,b), corresponds to a wavevector of 1.4~rad/$\mu$m, which leads to an excitation efficiency of $\mathrm{sinc}(kw_\mathrm{ant}/2)^2 = 0.998$, close to 1, i.e. the value for an antenna of vanishing width, Eq.~\ref{IdealizedAntenna}. This result shows that the analytical formula, Eq.~\ref{eq:e(t)atk0}, provides a solid basis to obtain design rules for SW demultiplexing. }

\subsection{Space-time evolution of wavepackets in absence of dissipation} 
We conclude that Eq.~\ref{eq:e(t)atk0} is sufficiently accurate to be used to illustrate how propagating wavepackets evolve along the magnetic stripe. 
Figure~\ref{fig:ModelVSAnal}(c) plots several such $\big|s^{k_0}_{21}(t,r)\big|/\big|s^{k_0}_{21}(0,0)\big|$. The calculations are now done in the zero damping limit to visualize the sole influence of dispersion on the amplitude decay. 

As expected, the arrival times scale with the propagation distance. The wavepackets broaden upon propagation as a result of dispersion: this spread of the pulse energy over a wider temporal window comes together with a reduction of the peak intensity. This broadening effect is more pronounced for the high frequencies due to the higher curvature of the dispersion relation (the higher \textit{"dispersivity"}). Noticeably, the rate of wavepacket broadening is slower than the rate of pulse separation. Figure~\ref{fig:ModelVSAnal}(c) evidences that pulses of duration $\tau_0=200$ ps carrying frequencies of 7, 10.5 and 15~GHz should be separable for sufficiently long travelling distances, at least according to the approximate analytical model.

\textcolor{black}{Therefore, the wave-packet amplitude decreases during propagation because of dispersion-induced broadening. This effect is captured by $C(r)$, which reduces the SW amplitude as the group-delay dispersion increases. This effect adds to the dissipation-induced decay and is non-negligible. The two contributions scale differently with the propagation distance $r$, see Eq.\ref{eq:e(t)atk0}. Table~\ref{tab:decay}, obtained for the parameters of Fig.~\ref{fig:ModelVSAnal}.c, shows that both contributions are comparable up to $r \approx 40~\mu$m, while Gilbert damping dominates at larger distances.
\begin{table}[h]
\caption{Normalized SW amplitude as a function of the propagation distance $r$, separating the contributions of dissipation-induced decay  and dispersion-induced broadening.}
\label{tab:decay}
\begin{center}
\begin{tabular}{cccc}
\toprule
$r$ [$\mu$m] & Damping $e^{-r/L_\text{att}}$ & Dispersion $C(r)/C(0)$ & Combined \\
\midrule
20  & 78\,\% & 81\,\% & 63\,\% \\
40  & 61\,\% & 65\,\% & 40\,\% \\
60  & 47\,\% & 54\,\% & 25\,\% \\
80  & 37\,\% & 47\,\% & 17\,\% \\
100 & 29\,\% & 41\,\% & 12\,\% \\
\bottomrule
\end{tabular}
\end{center}
\end{table}
}

\subsection{Evolution of the wavepacket integrity upon propagation} 
\begin{figure}
	\includegraphics[width=0.45\textwidth]{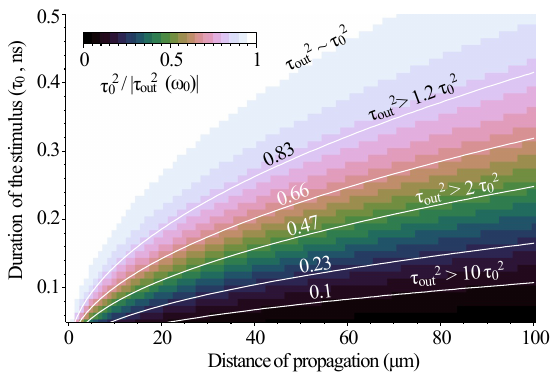}
	\caption{Temporal spreading $\tau_0^2 / \|\tau_{\mathrm{out}}^2\|$  of spin-wave wave packets as a function of the excitation duration $\tau_0$ and propagating distance. The parameters are similar to Fig.~\ref{fig:ModelVSAnal}. 
    }
	\label{fig:SpreadingVStau}
\end{figure}

The temporal spreading of wave packets depends on their initial excitation durations $\tau_0$ and varies asymptotically like $\propto \sqrt r$ for large propagation distances (see Eq.~\ref{DefinitionTauOut}). The increase of the wave packet duration can be measured by the \textit{"integrity"} figure of merit $\tau_0^2 / \|\tau_{\mathrm{out}}^2\| \in [0,~100\%]$, formed the ratio between the input and output temporal widths of the wavepackets. 
Fig.~\ref{fig:SpreadingVStau} reports a quantitative evaluation of this metric for the example of a 10~GHz pulse, now with dissipation accounted for. 

For long pulses, the integrity of the wavepacket is maintained (i.e. $\tau_0 \approx |\tau_{\mathrm{out}}|$, white area in Fig.~\ref{fig:SpreadingVStau}) up to large propagation distances. In contrast, for short pulses ($\tau_0 \ll0.4$~ns), the wavepacket rapidly broadens with the propagation distance (coloured areas in Fig.~\ref{fig:SpreadingVStau}). In this regime the output temporal width is no longer solely set by the excitation duration but rather more and more by the dispersive nature of the SWs as they propagate. 

\section{Frequency demultiplexing using spin waves}	

	\begin{figure}
    \centering
		\includegraphics[width=0.49\textwidth]{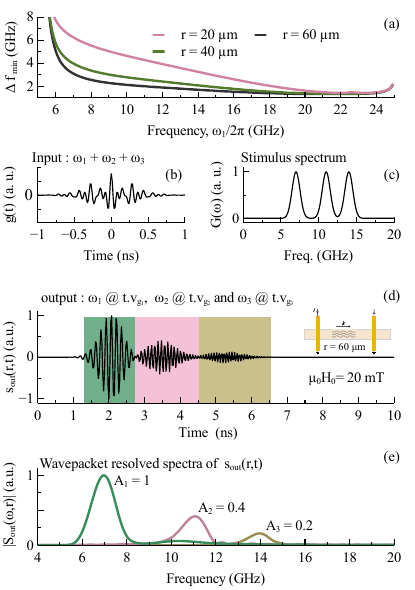}
		\caption{Determination of the separable frequency pairs and demonstration of spatio-temporal demultiplexing. (a) Relation between propagation distance and minimum frequency spacing $\Delta f_{\mathrm{min}}$ (Eq.~\ref{eq:criteria_tau_in}) required to separate two wavepackets for $\tau_0 = 0.3$~ns. (b) Complex field pulse of duration $\tau_0 = 0.3$~ns carrying the frequencies $\{f_1,f_2,f_3\} = \{7, 11, 14\}$~GHz. (c) Positive-frequency part of the corresponding spectrum. (d) Time-resolved voltage expected at a distance $r=60~\mu$m for single-wire antennas (see their sketch in inset). The coloured backgrounds indicate the temporal gates used for time gating. (e) Power spectral density of these time gated signals.}
		\label{fig:criteria}
	\end{figure}
\subsection{Criterion for the separability of wavepackets}
To achieve proper wavepacket separation, the temporal spacing between the centers of two wavepackets $\Delta\tau_g(\omega_1,\omega_2)$ should be larger than the sum of their widths. This is written as:		
\begin{equation}
	\Big|\Delta\tau_g(\omega_1,\omega_2)\Big|
	\;\geq\;\\ \frac{1.18}{\tau_0}
    \left[||\tau_\textrm{out}^2(\omega_1)|| +||\tau_\textrm{out}^2(\omega_2)\|\right]
	\label{eq:criteria_tau_in}
\end{equation}
and this sets conditions on the frequencies $f_1$ and $f_2>f_1$. \textcolor{black}{Note that the factor 1.18 corresponds to the half width at half maxima of the wavepackets. A larger factor leads to a more robust separation, at the cost of a lower spectral resolution (see Supplementary Material).} Figure~\ref{fig:criteria}.a shows the minimum frequency spacing $\Delta f_{\mathrm{min}} = ({\omega_2 - \omega_1})/({2\pi})$ required to separate 300 ps-long wave packets, as calculated using the analytical framework. For instance, a wave packet propagating over a distance $r=60\,\mu$m carrying a frequency $f_1 = 11$~GHz can be separated from any other carrying $f_2 > 13$~GHz. Since the separability criterion (Eq.~ \ref{eq:criteria_tau_in}) assumes an infinitely narrow antenna and a perfectly parabolic dispersion relation, we must validate it with the results of the full numerical integrations using Eq.~\ref{eq:Eout}.

\subsection{Demonstration of the demultiplexing of three frequencies}
\label{subsection: Demonstration of the demultiplexing}
 From Fig.~\ref{fig:criteria}.(a) we have identified  $\{f_1, f_2, f_3\} = \{7, 11, 14\}$~GHz with respective wavevectors $\{k_1, k_2, k_3\} = \{0.53, 1.62, 2.95\}\,\textrm{rad}/\mu$m, as carriers that should be separable for a pulse duration of 300 ps and antennas of zero widths. This stems from their differing group velocities which are $\{v_{g1}, v_{g2}, v_{g3}\} = \{29.8, 17.5, 11.6\}\,\textrm{km/s}$. This section demonstrates this statement for antennas of widths 100 nm by fully accounting for dispersion and dissipation using exact integration (Eq.~\ref{eq:Eout}). \textcolor{black}{It is clear that near the group-velocity peak (${\sim}$6--8~GHz), $\Delta f_\text{min}$ is diverging, indicating that demultiplexing operations are not achievable for these frequencies.
 }
 
 Fig.~\ref{fig:criteria}.(b,c) display the sum of the three field pulses in time and frequency domains, as applied on the first antenna. The goal is to separate the three pulses. Fig.~\ref{fig:criteria}.(d) reports the signal expected on the receiving antenna placed at a distance $r=60~\mu$m. 
Three successive wavepackets contribute to the signal. The first arriving wavepacket ($t \in [0,2.7]$ ns) corresponds to the SWs with the highest group velocities; this packet carries a pseudo-period corresponding to the lowest frequency component ($f_1 = 7$~GHz). The two subsequent wavepackets, observed in the 2.6--4.5~ns and 4.5--6.5~ns time intervals, display lower amplitudes and their pseudo-period correspond to excitation frequencies of 11 and 14~GHz, respectively. Temporal gates (shaded regions in Fig.~\ref{fig:criteria}.d) can be applied to analyse the spectral contents of each wavepacket, displayed in Fig.~\ref{fig:criteria}.e. 
The gated spectra confirm three wavepacket assignments to the three frequencies initially injected into the system. This demonstrates the demultiplexing of the three initially overlapping wave packets.
\begin{figure}
\centering
	\includegraphics[width=0.49\textwidth]{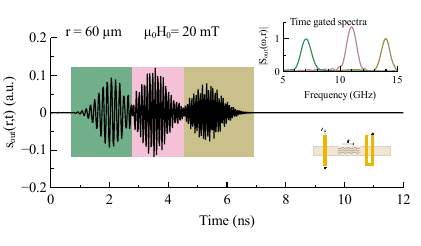}
	\caption{ Time-resolved voltage expected at a distance $r=60~\mu$m for a single wire emitting antenna and a $\sqcup$-shaped receiving antenna (sketch in bottom inset). The coloured backgrounds indicate the temporal gates used for time gating. Top inset: Normalized power spectral density of these time gated signals.}
	\label{fig:Demux_New_set_ant}
\end{figure}
\subsection{Antenna engineering for amplitude-conserving demultiplexing}
The successive wavepackets have unfortunately different amplitudes, $A_1$, $A_2/A_1 =0.4$ and $A_3/A_1 =0.2$. This is arising since arriving later in time means a further amplitude decay through Gilbert damping and dispersion broadening. 

This non-conservation of the amplitudes of the 3 stimuli is a clear drawback. Fortunately, it can be compensated by a proper design of the antennas, either by \textit{pre-accentuating} the wavepackets at their generation and/or by enhancing their detection yield at the receiving antenna. The objective is to design two antennas such that 
\begin{equation} || \tilde h{_\textrm{rf}}^\textrm{ant1}(k_i).  \tilde h{_\textrm{rf}}^\textrm{ant2}(k_i)^* || \propto (A_i)^{-1} ~\textrm{for} ~i=1,2,3 \label{amplitudeConcervationCriterion}\end{equation} while maximizing $||\tilde h_\textrm{rf}(k_3)||^2$.
These three criteria cannot be jointly fulfilled with the single-wire antenna used so far, which has $w_\textrm{ant}$ as single design parameter.

The simplest design is to use a single wire antenna of small width $L_1$ as a broadband emitter and a $\sqcup$-shaped antenna as a filtering receiver. The latter is formed by two wires of widths $L_1$ carrying opposite currents, placed at wire-to-wire  center-to-center distance $c$. Assuming infinitely thin antennas, the $x$-component of the efficiency function becomes: 
\begin{equation}
h_{rf}^S(k_x)  h_{rf}^\sqcup(k_x)^*=\textrm{sinc}\left(\frac{k_xL_1}{2}\right)^2\times \textrm{sin}\left(\frac{k_xc}{2}\right).
\label{eq:U-shape + Single}
\end{equation}
Using Eq.~\ref{eq:U-shape + Single} we determine the pairs $\{L_1, c\}$ that ensure the proper gradation of antenna efficiencies (eq.~\ref{amplitudeConcervationCriterion}) at the  wavevectors $\{k_1, k_2, k_3\} = \{0.53, 1.62, 2.95\}\,\textrm{rad}/\mu$m that correspond to our three frequencies $\{f_1, f_2, f_3\} = \{7, 11, 14\}$~GHz. These criteria are reasonably satisfied for the pair $\{L_1=100~\textrm{nm}, c=200~\textrm{nm}\}$.

We performed the numerical integration to test this combination of antennas. The results are gathered in Fig.~\ref{fig:Demux_New_set_ant}. The time-domain voltages and their spectra indicate that this $\mathrm S + \sqcup$ design successfully implements frequency demultiplexing which a much improved conservation of the amplitudes of the wavepackets. Still, the voltage induced by the $f_2$ wavepacket is slightly higher than the two others. This arises because the $\mathrm S + \sqcup$ antenna design lead to a slightly too high $||\tilde h_\textrm{rf}(k_2)||^2$. This could be solved by allowing additional degrees of freedom in the antenna design. This versatility allows to design amplitude-conserving frequency demultiplexing operations for arbitrary sets of frequencies.


\section{Conclusion}

We have described the spatio-temporal propagation of spin-wave wavepackets evolving in magnetic stripes, as driven by pulses of microwave magnetic field. We have clarified the influences of the excitation duration and of the spin wave dispersion relation, as well as the influence of the spin-wave linewidth.  This led us to propose a criterion for the spatio-temporal separation of initially overlapping wavepackets that we illustrate by demultiplexing a complex signal that comprises three distinct frequencies, while preserving their relative amplitudes.



\textcolor{black}{
As all other SW-based devices, our multiplexer has a limited operational bandwidth that is set by the spin-wave dispersion relation. Our proposed concept is therefore limited to a few tens of GHz, essentially determined by the FMR frequency, the antenna shape and width, and the bias field required to obtain the desired FMR frequency and magnetization direction. } 

\textcolor{black}{For the proposed device to be all-electrical and therefore suitable for full integration, the difficulty will be the on-chip generation of the field needed to bias the magnetization in the direction transverse to the spin wave conduit. Replacing an external magnet by exchange biasing over- and under-layers may be a solution; this however calls for material research on the generation of strong exchange bias fields that would preserve the Gilbert damping of the ferromagnetic layer.}
Future work will leverage this principle to design devices capable of filtering and processing multiple spin-wave channels in parallel.

\begin{acknowledgments}
This work was supported by the French RENATECH network, by the French National Research Agency (ANR) as part of the “Investissements d’Avenir” and France 2030 programs. This includes the PEPR SPIN projects ANR 22 EXSP 0008 and ANR 22 EXSP 0004. We thank Claude Chappert and Loukas Kokkinos for helpful comments.
\end{acknowledgments}

\section*{AUTHOR DECLARATIONS}
\subsection*{Conflict of Interest}
The authors have no conflicts to disclose.

\subsection*{Author Contributions}
\textbf{G.~Y.~Thiancourt}: Data curation (lead); Formal analysis (lead); Writing– original draft (equal); Writing– review and editing (equal); Investigation (equal).
\textbf{T.~Devolder}: Funding acquisition (lead); Investigation (equal); Writing– original draft (equal); Writing– review and editing (equal); 

\section*{DATA AVAILABILITY}
The data that support the findings of this article are openly available \cite{MyZenodo}.

\input{apssamp.bbl}
\newpage
\begin{widetext}
\begin{center}
\textbf{\large Supplemental Materials: Modelling of spin-wave wavepackets excited by an RF burst for signal demultiplexing}
\end{center}
\end {widetext}
\textit{Abstract: This is supplementary of the paper "Modelling of spin-wave wavepackets excited by an RF burst for signal demultiplexing" by G. T. Thiancourt and T. Devolder.}

\renewcommand{\theequation}{S\arabic{equation}}
\renewcommand{\thefigure}{S\arabic{figure}}
\renewcommand{\bibnumfmt}[1]{[S#1]}
\renewcommand{\citenumfont}[1]{S#1}
\setcounter{figure}{0}
\setcounter{section}{0}

\section{Hysteresis}
\label{app:Hysteresis}
\begin{figure}[h!]
    \centering
    \includegraphics[width=1\linewidth]{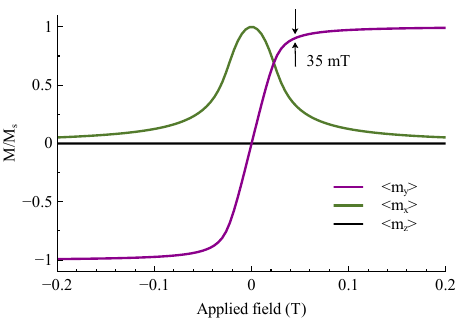}
    \caption{Hysteresis loop of a CoFeB waveguide with a thickness of 60~nm and a width of 5~\textmu m for a transverse applied field obtained from micromagnetic simulations using TetraX2 with a spatial discretization of $4 \times 4$~nm$^2$. }

    \label{fig:hysteresis}
\end{figure}

Fig.~\ref{fig:hysteresis} displays the hysteresis loop calculated using TetraX2 \cite{korber_finite_2021}. For the applied magnetic field of 35~mT used throughout this work, the waveguide is close to saturation. One can thus expect that the analytical expression given in Eq.~\ref{wk} can be employed. We have validated this assumption by eigenmode simulations in Fig.~\ref{fig:dispersion_relation}.c.

\section{Linearized susceptibility}
\label{app:deltaomega0}
Implementing Eq.~\ref{eq: S21} requires the evaluation of the susceptibility. For this purpose we still consider a thin ferromagnetic stripe of thickness $t_{\mathrm{mag}}$, width
$w_{\mathrm{stripe}}$, and translational invariance along $x$. To get an analytical expression of the susceptibility we disregard the edge domains and assume a uniform internal static field $\mathbf{H}_0 = H_0 \hat{\mathbf{y}}$ that leads to a uniform equilibrium magnetization $\mathbf{M}_0 = M_s \hat{\mathbf{y}}$.

Small-amplitude spin wave (SW) are excited by a transverse rf magnetic field
$\mathbf{h} = (h_x,0,h_z)\,e^{i(\mathbf{k}\cdot\mathbf{r}-\omega t)}$, inducing
magnetization fluctuations
$\mathbf{m} = (m_x,0,m_z)\,e^{i(\mathbf{k}\cdot\mathbf{r}-\omega t)}$, with
$\mathbf{k} = (k_x,k_y,0)$ and
$k^2 = k_x^2 + n^2\pi^2/w_{\mathrm{stripe}}^2$.

Linearizing the Landau--Lifshitz--Gilbert equation in the weak-damping limit
($\alpha \ll 1$) yields the coupled equations for the transverse components
$(m_x,m_z)$,
\begin{equation}
	\begin{pmatrix}
		\Omega_x(k) - i\alpha\omega & -i\omega \\
		+i\omega & \Omega_z(k) - i\alpha\omega
	\end{pmatrix}
	\begin{pmatrix}
		m_x \\ m_z
	\end{pmatrix}
	=
	\omega_M
	\begin{pmatrix}
		h_x \\ h_z
	\end{pmatrix},
\end{equation}
where $\omega_M = \gamma\mu_0 M_s$ and the effective stiffness terms read
\begin{align}
	\Omega_z(k) &= \omega_H + \omega_M(1-P), \\
	\Omega_x(k) &= \omega_H + \omega_M P\,\frac{k_x^2}{k^2},
\end{align}
with $\omega_H = \gamma\mu_0 H_0$ and $P$ the classical \cite{herring_theory_1951} demagnetizing factor expressed in reciprocal space as
\begin{equation}
	P(k_n t_{\mathrm{mag}})
	=
	1 + \frac{e^{-k t_{\mathrm{mag}}}-1}{k t_{\mathrm{mag}}}.
\end{equation}
The magnetic susceptibility tensor
$\boldsymbol{\chi}(k,\omega)$ is obtained by inverting the above matrix. In the
transverse $(x,z)$ basis, its non-zero components read
\begin{align}
	\chi_{xx}(k,\omega) &= \frac{\omega_M\,\Omega_z(k)}{\Delta_\alpha(k,\omega)}, \\
	\chi_{zz}(k,\omega) &= \frac{\omega_M\,\Omega_x(k)}{\Delta_\alpha(k,\omega)}, 
\end{align}
where all components share the common denominator
\begin{equation}
	\Delta_\alpha(k,\omega)
	=
	\Omega_x(k)\Omega_z(k)
	-
	\omega^2
	-
	i\alpha\omega\big[\Omega_x(k)+\Omega_z(k)\big].
\end{equation}
Since we have assumed a uniform equilibrium magnetization $\mathbf{M}_0 \parallel \hat{\mathbf{y}}$, the longitudinal susceptibility $\chi_{yy}$ vanishes.

\section{Derivation of the phase shift to second order in $k$}
\label{app:analytical}

We make a series of approximations in the algebra leading to Eq.~\ref{eq:e(t)atk0}. First, we treat the antenna as infinitely narrow, following Eq.~\ref{IdealizedAntenna}. Second, we expand the magnetic susceptibility to second order in wavevector to capture the curvature of the dispersion relation; the validity of this approximation is assessed in Section~\ref{ComparaisonDesModeles}.
We rewrite the frequency near $\omega_0$ by setting $\omega=\omega_0+\delta\omega$. Expanding the dispersion relation to second order around $k_0$, with $\delta k=k-k_0$, and assuming $\omega_0\gg v_g\,\delta k+(\tau_D^2v_g^3\,\delta k^2)/r+\delta\omega$, the magnetic susceptibility becomes
\begin{equation}
	\tilde\chi^{xx}(k_x,\omega)=\frac{\omega_\textrm{max}(k_0)^2}{2\omega_0\left(v_g\,\delta k+(\tau_D^2v_g^3\,\delta k^2)/r+\delta\omega\right)-i\omega_0\Delta\omega(k_0)},
    \label{eq:ChiInSupplementary}
\end{equation}
where $\omega_\textrm{max}^2(k_0)=\gamma_0^2 M_s\left(H_0 + M_s\left(1-P(k_0)\right)\right)$ and $P(k_0)$ is the simplified dipolar factor. Integrating $\tilde S_{21}(\omega,r)$ around $k_0$ yields
\begin{equation}
	\tilde S_{21}(\omega,r)\propto\frac{i\,\omega_\textrm{max}^2(k_0)}{v_g\,\omega_0}\,h_{rf}(k_0)^2\,e^{-\frac{\Delta\omega_0\, r}{2v_g}}\,e^{-ik(\omega)r}.
\end{equation}
Neglecting slow terms (i.e.\ $r\,\Delta\omega_0\,\tau_D\,\delta\omega$), the phasor can be rewritten as
\begin{equation}
 	  ik(\omega)r=i\left(k_0 r+\tau_g\,\delta\omega+\tfrac{1}{2}\tau_D^2\,\delta\omega^2+
    \mathcal{O}(\delta\omega^3)\right).   
\end{equation}
We thus recover the second-order expansion of the phase shift in $\delta\omega$ (Eq.~\ref{eq:Phi_expand}).

\section{Phase oscillation within the pulse envelope} 
The phasor term $\cos[\varphi(r,t)]$ in eq.~\ref{eq:phase} captures both the wave character of the carrier and the dispersion-induced quadratic phase (chirp). Its full expression is:
\begin{equation}
\begin{split}
\varphi(r,t)
=
\underbrace{
-\frac{\tau_D^2 (t-\tau_{\mathrm{vg}})^2}{2\tau_{\mathrm{out}}^4}
-\frac{1}{2}\tan^{-1}\!\left(\frac{\tau_D^2}{\tau_0^2}\right)
}_{\text{chirp-induced phase}}
\\\;+\;
\underbrace{(k_0 r - \omega_0 t)}_{\text{carrier wave}}
\end{split}
\label{eq:phaseInSupplementary}
\end{equation}
The carrier term describes the underlying harmonic oscillation with a phase front advancing at $r=\omega_0 t / k_0$. The quadratic contribution in eq.~\ref{eq:phase} represents a time-dependent chirp centred around $r=v_g t$. It reflects a continuous redistribution of instantaneous frequencies across the pulse, with different temporal components acquiring distinct phase shifts during propagation. The strength of this chirp is governed by the competition between the dispersion timescale $\tau_D^2$ and the effective temporal width $\tau_\textrm{out}^4$, leading to a progressive deformation of the phase profile as the wavepacket broadens.
The Arctan term corresponds to a global phase offset that depends solely on the ratio $\tau_D/\tau_0$. This contribution is analogous to a Gouy-like phase shift \cite{Feng_Physical_2001,fujita_encoding_2017,jia_chiral_2021}: it does not modify the local frequency content of the wavepacket --- around $t=r/v_g $ --- but represents a cumulative phase lag acquired during propagation. Physically, it originates from the same dispersive mechanism responsible for temporal broadening and can be interpreted as the phase counterpart of the energy spreading in time. It is worth noting that the instantaneous angular frequency, obtained by differentiating Eq.~\ref{eq:phase}, grows without bound as $t\to \infty$, which could seem unpysical at first sight. However, this divergence is not happening since the phasor factor is weighted by a Gaussian envelope that suppresses all contributions at $|t-\tau_\text{vg}| \gtrsim \tau_\text{out}$. The phase expression is therefore only meaningful within the temporal support of the pulse, where the chirp remains a small correction to the carrier frequency $\omega_0$.

\section{Case of the vanishing group velocity} 
Let us examine the flat-band limit when $v_g \to 0$. This pathological situation lies outside the validity range of the Eq.~\ref{eq:e(t)atk0} since the expansion in Eq.~\ref{eq:ChiInSupplementary} required a well-defined group velocity. 

For $v_g \to 0$ we need to restart from eq.~\ref{eq: S21}. The dispersion reduces to $\omega(k_x)=\omega_\textrm{FMR}$ with a constant linewidth such that the susceptibility is independent from the wavevector. It is $\tilde \chi^{xx}(k_x,\omega)=\tilde \chi^{xx}(k=0,\omega) \equiv \tilde \chi^{xx}(\omega)$, as if there was a unique SW and if it was confined in all directions. Using eq.~\ref{eq: S21} the transmission parameter is then simply: 
\begin{equation}
    \tilde S_{21}(\omega,r)=\tilde \chi^{xx}(\omega)\,\delta_{r=0}, \label{S21vanishingVg}
\end{equation} 
For $r>0$ the transit time $r/v_g$ diverges; the wavepacket spends an infinite time in the absorbing medium and is completely attenuated when arriving at the position $r$. There is simply no signal at the receiving antenna, as expressed from the Dirac function in eq.~\ref{S21vanishingVg}. 
\\For $r=0$ we recover the classical VNA-FMR configuration \cite{bilzer_vector_2007} when a single antenna at $r=0$ is used for both emitting and receiving the SWs, and the $\tilde S_{21}(\omega,r)$ signal is directly linked with the susceptibility and only present below the emitting antenna.

The corresponding impulse response $s_{21}(t,r)$ is then simply:
\begin{equation}
    s_{21}(t,r)\propto\left(\sin(\omega_\textrm{FMR}t)\exp(\frac{-|t|\Delta\omega_\textrm{FMR}}{2})\right)\delta_{r=0}.
\end{equation}

The transmission response to a Gaussian stimuli $\{g(t),\,\tilde G(\omega)\}$ is, therefore, defined as a multiplication in frequency domain,
\begin{equation}
    \tilde S_\textrm{out}(\omega,r)=\left(\tilde G(\omega)\, \tilde \chi^{xx}(\omega)\right)\delta_{r=0},
\end{equation}
or equivalently the time-domain convolution:
\begin{equation}
    s_{\mathrm{out}}^{v_g=0}(t,r)= \left(s_{21}(t,r)\, \ast g(t) \right)\delta_{r=0}.
\end{equation}

\begin{figure}
   \centering
    \includegraphics[width=1\linewidth]{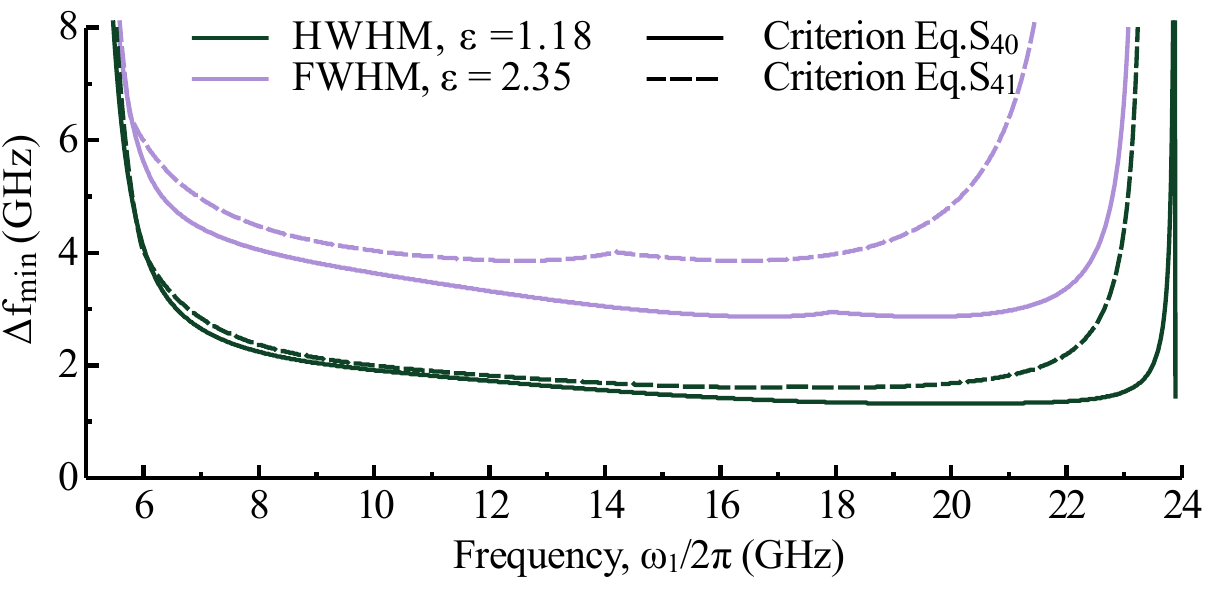}
    \caption{Antenna efficiency function for a 100 nm wide stripline antenna directly deposited on the magnetic surface.}
    \label{fig_sup:criteria}
\end{figure}

\textcolor{black}{\section{Other criteria for the separability}
In general, the  criterion for the separability of two wavepackets can be parametrized by a positive number $\epsilon$. This reads:
\begin{equation}
	\left|\Delta\tau_g(\omega_1,\omega_2)\right| \geq \frac{\epsilon}{\tau_0}
	\left[\left\lVert\tau_{\textrm{out}}^2(\omega_1)\right\rVert
	+ \left\lVert\tau_{\textrm{out}}^2(\omega_2)\right\rVert\right].
	\label{eq:criteria_epsilon}
\end{equation}
In the main document, we have chosen $\epsilon=1.18$ to have wavepackets separated by the quadratic sum of their half width at half maximum (HWHM). $\epsilon=2.35$ would account for a separation set by the full widths at half maximum (FWHM) of the two wavepackets.}

\textcolor{black}{An alternative separability ansatz is the use of a Rayleigh-like criterion, where the spacing between the two wavepackets must exceed the temporal width of each of them; this reads:
\begin{equation}
	\left|\Delta\tau_g(\omega_1,\omega_2)\right| \geq \frac{2\epsilon}{\tau_0}
\max\left[\left\lVert\tau_{\textrm{out}}^2(\omega_1)\right\rVert,\,
\left\lVert\tau_{\textrm{out}}^2(\omega_2)\right\rVert\right].
	\label{eq:criteria_Rayleigh}
\end{equation}}

\textcolor{black}{Fig.~\ref{fig_sup:criteria} compares the two criteria for $\epsilon=1.18$ and $\epsilon=2.35$. As expected, tightening the separability criteria by increasing $\epsilon$ raises the required spacing of the two carriers $\omega_1$ and $\omega_1$. At low frequencies, the two criteria (eqs.~\ref{eq:criteria_epsilon} and \ref{eq:criteria_epsilon}) yield similar constraints. In contrast, when aiming at higher carrier frequency $\omega_1$, the Rayleigh-like criterion yields a more demanding (larger) $\Delta f_\textrm{min}$ than Eq.~\eqref{eq:criteria_epsilon}. This is due to the difference of the width of the two wavepackets, which tightens the Rayleigh-type criterion.}

\end{document}

%% file: apssamp.bbl
%